# Preferential Detachment in Broadcast Signaling Networks: Connectivity and Cost Trade-off

May Lim[1,2,*], Dan Braha[1,3] Sanith Wijesinghe[1], Stephenson Tucker,[4] and Yaneer Bar-Yam[1]

[1]New England Complex Systems Institute
[2]Brandeis University
[3]University of Massachusetts Dartmouth
[4]Sandia National Laboratory

## Abstract

We consider a network of nodes distributed in physical space without physical links communicating through message broadcasting over specified distances. Typically, communication using smaller distances is desirable due to savings in energy or other resources. We introduce a network formation mechanism to enable reducing the distances while retaining connectivity. Nodes, which initially transmit signals at a prespecified maximum distance, subject links to preferential detachment by autonomously decreasing their transmission radii while satisfying conditions of zero communication loss and fixed maximum node-hopping distance for signaling. Applied to networks with various spatial topologies, we find cost reductions as high as 90% over networks that are restricted to have all nodes with equal transmission distance.

*M.L. is on leave from the National Institute of Physics, University of the Philippines.

# I. Introduction

The understanding of complex networks with physical links has been steadily improving over the years [1] yielding insights into network robustness to failure and attack [2-6], navigability [7], jamming and congestion [8-12], speed of information propagation [13], and efficient routing [14-20]. In this paper we focus our attention on networks in which physical links are absent but connections are possible through spatial signaling by nodes using broadcasting of message carriers (e.g. radio wave, chemical) into the space around them. In such broadcast communication networks, where a physical link is absent between nodes, internode communication is established through virtual links which are created when a node sends out a signal that is received by all nearby nodes. Among a large number of such biological networks using chemical signaling, paracrine signaling is an example of a neuronal molecular signaling mechanism that involves the secretion of chemical signals onto a group of nearby target cells [21]. Broadcast signaling is also used in electromagnetic transmission in multihop wireless networks, where messages may traverse multiple wireless links in order to reach a destination [22]. In both cases, the localized broadcast nature of transmission imposes physical constraints on the system. First, all nodes within broadcast range of a transmitting node receive the signal, and only these nodes receive the signal. If signaling is not perfect, probabilistic distance-dependent link formation can be considered [23]. Second, cyclical retransmission of the same signal is avoided using refractory periods or other methods. Third, quite generally, there is a signal lifetime or equivalently a maximum number of signal retransmission hops through intermediate nodes. Such built-in lifetimes perform the dual function of discarding stale information and preventing system overload. Fourth, such networks generally expend energy and other resources to transmit signals, with a cost that increases with the distance of transmission, a key consideration in the effective design of networks, whether chemical or radio. Finally, a node that receives the same type of signal (i.e. over the same frequency or with the same chemical) from two different nodes at the same time may have difficulty decoding the signals correctly if they use the same signaling channel. When the occurrence of such events become significant (e.g. high signal transmission rates in high node density regions), signal aggregation or conflict avoidance and error resolution techniques are necessary to address the problem. A thorough analysis of such smart protocols [24] is beyond the scope of our current work. To avoid this problem biological networks, such as the neural network of the brain, use a large number of distinct chemicals to enable multiple channels [25, 26], while radio networks most commonly use labeled signal packets.

When barriers and other physical constraints effectively fix the spatial location of nodes and connectivity (whether direct or mediated) is required for all node pairs, the transmission distance of each node is the only remaining variable of the system. Since the cost in energy or chemical messenger to establish links grows as $D_i^\alpha$, where $2 \leq \alpha \leq 4$ for electromagnetic waves as well as chemical signals, the best cost reduction scheme will come from an effective reduction of the transmission distance. Truly autonomous wireless nodes are self-powered, typically by batteries. Biological cells have limited stores of available energy and both energy and material considerations are key concerns in biological systems. In either case reducing the distance of transmission to reduce costs is important.

In the following we consider a network of point nodes transmitting and receiving signals. This may be considered a model of multihop wireless networks [27-31], or of biological cells with chemical signaling. The underlying assumption of much of the earlier work in spatial networks is that nodes are randomly distributed [32] or node location can be controlled [22]. In many systems, however, node spatial distribution is constrained, e.g. by the means by which nodes are distributed, and cannot always be independently controllable. As such, widely-known methods of preferential attachment [33] and link rewiring [34] are not readily applicable. Instead, we propose a simple "preferential detachment" algorithm that minimizes the number of links in high-density areas while maintaining links in low-density areas.

Unlike continuum percolation [35] and other local information-based generative network models [36], which emphasize growth by establishing links, our method stresses the role of pruning. Synaptic overgrowth and pruning [37] is a standard process for changing networks during the maturation of the mammalian neural system [38]. In order to evaluate the benefits of our pruning-based, preferential detachment algorithm for network formation we compare its performance with uniform transmission distance networks for various node topologies.

We find that preferential detachment results in significant cost reduction, saving 80-90% for a wide variety of tested network topologies. It also results in the reduction of signal congestion when we consider the message retransmission. Thus, our work, while distinct in its specific approach, can also be considered as a congestion reduction algorithm, called "congestion aware" protocols [6, 11, 14-17, 20].

**II. Network Model**

We distribute $N$ nodes in a square $s \times s$ grid where the $i$th node broadcasts over a transmission distance $D_i$ from its location, $(x_i, y_i)$. If another node $j$ is at a distance $d_{ij} \leq D_i$, a link is established from $i$ to $j$ with a path length $L_{ij} = 1$ from $i$ to $j$. For the general case of variable $D_i$, a link from $i$ to $j$ does not imply a link from $j$ to $i$. Upon receiving a signal, a node retransmits the signal up to a maximum of $h$ links. Furthermore, we assume functionally identical nodes that control only their own transmission power. We perform simulations on networks having different topologies all with $N = 256$ and $s = 600$ and then characterize the scaling behavior of the network properties to larger network sizes.

We characterize each $h$–constrained network structure of every investigated topology using: (1) the reachable pairs fraction $R = n/N(N-1)$, where $n$ is the number of distinct ordered pairs $(i, j)$ such that $i$ can transmit to $j$ through less than $h$ links; (2) the average path length $L = \Sigma L_{ij}/n$ of all reachable pairs; and (3) a normalized network cost $C = (\Sigma D_i^2)/D_o^2$, where $D_o^2$ is the cost for a node at the center of the grid to broadcast over the entire area directly ($D_o$ is half the grid diagonal length $= s\sqrt{2}/2$ and $L_{oj} = 1$ for all $j$) where the scaling of cost with distance is $\alpha = 2$. We use $\alpha = 2$ as the reference case, which is the power needed in free space, or the amount of chemical signal needed with a chemical wave propagation front. ($\alpha$ increases for other assumptions of power or chemical spreading or dissipation with distance). Our analysis of the benefits of reduction of $D_i$ is then conservative as the higher values $\alpha > 2$ yield greater cost reductions.

**III. Uniform distance networks**

Uniform transmission distance networks, where $D_i = D$ for all nodes $i$, have previously been studied as a percolation problem in a two-dimensional random lattice in which bonds are determined by the distance between sites [39]. In such a percolation problem, $N$ sites are randomly located in an $s$ x $s$ square grid. When the distance between two sites is less than $r_s$, a bond is formed between them. For $r_s$ greater than the critical (percolation) radius $r_s = r_{sp}$, one can find a series of links that traverses the space in each linear dimension. Monte Carlo simulations have shown that $r_{sp} = (1.06 \pm 0.03)\, 2s/\sqrt{(\pi N)}$ [39]. The analytic solution for $N$ nodes in a disc of unit area is given by $\pi r_{sp}^2 = (\log(N) + c(N))/N$, which yields a fully connected network only in the asymptotic limit $c(N) \to \infty$ [40]. We can use these results to infer properties of our network for random distributions of sites. For our uniform radius network, (Fig. 1b, $h = 20$ hops), we expect that the average path length $L$ reaches its peak value, $L(r_p \approx 50) \approx 10.5$ hops, near the critical radius of percolation. The critical radius should also be the value of most rapid change of the reachable pairs fraction $R$. If we do not limit the maximum number of hops $h$ (as we do in subsequent studies), $r_{sp}$ gives a lower bound on the node transmission radius for boundary-to-boundary connectivity, albeit in the presence of "dead spots" indicating isolated clusters ($R < 1$). Imposing substantially smaller $h$ necessarily increases all limiting radii values, thus $r_{sp}$ is a strict lower bound. For the node distribution in Fig. 1a, we calculate $r_{sp} = 45 \pm 1.3$, which is consistent with Fig. 1b. While percolation is concerned with widespread communication (which has known analytic solutions that converge to full connectivity $R = 1$ in large number limits), our concern is complete communication over a small number of nodes. The relationship between the percolation problem and our multihop network ends with the determination of the lower bound: our interest is in finding the minimum radius $r_{min} > r_p$ with full nodal coverage ($R = 1$) for a finite $N$, which is necessarily a higher value ($r_{min} = 60$ in Fig. 1a). In most cases, $r_{min}$ decreases monotonically with increasing $h$.

Figure 1a shows the uniform distance network with minimum radius that still is fully connected, and thus is at minimum cost, formed by $N = 256$ nodes distributed randomly over a 600 x 600 space with $D_i = D = 60$ (for all nodes $i$), $h = 20$ hops, and $C = ND^2/D_o^2 = 5.12$. Two concerns arise with uniform distance networks: (1) The existence of the $D$-determining proverbial "weak link" – network fragmentation immediately results when the node responsible for it is lost; and (2) Unnecessary transmission cost in regions of high node densities, i.e. $D \gg d_{ij}$. The first concern can be addressed by increasing the radius of transmission to be larger than the minimum distance. The solution to the latter is selectively reducing $D_i$ in high node density locations.

An apparently intuitive solution, using local density information for selectively reducing $D_i$, can be shown to be generally ineffective. Savings in regions of high density are often compensated for by unnecessary costs outside that region when a single functional form is used to adjust transmission for all nodes. This can be seen through direct analysis of a three node network using, e.g. $D_i = D_M /\rho_{local}$ [41]. More generally, setting the radius based upon local density is ineffective because the density is not isotropic, so that nodes that are near the edges of clusters defeat optimization by simple algorithms. Indeed, simulations show that a variety of density dependent rules do not improve significantly on uniform radius protocols [41]. We are thus motivated to find an adaptive method that overcomes this limitation.

**IV. Preferential detachment**

In our preferential detachment algorithm, we consider adaptive adjustment of node transmission distance using an algorithm that reduces $D_i$ until a minimum criterion for effective transmission is breached. In addition to complete network connectivity, we further restrict the networks to strictly observe maximum number of allowed hops $h$. The adaptive distance network formation begins with establishing the best uniform distance network by setting the uniform radius to some maximum value (typically of the order of $s$) and performing synchronous radii reduction. When the network connectivity breaks ($R < 1$), we incrementally increase the radius so that the network is fully connected. The adjustment of individual node radii then occurs as an asynchronous update of each node according to the following protocol:

1. Node *i* broadcasts a signal to all nodes. Nodes receiving the signal respond to the initial request with a response that can be detected by the original transmitter as a confirmation of receipt. Signal retransmission is allowed until $h$ is reached.

2. Node *i* decrements its transmission distance by a fixed amount ($D_i' = D_i - \delta$).

3. Node *i* resends a signal.

4. If node *i* receives the same number of receipt replies, it returns to step 2. Otherwise, the node fixes its transmission strength.

We note that the sender need only measure the aggregate magnitude of the response signal not identify each of the signals separately, which enables a wider range of application contexts including chemical and wireless signals. The cycle repeats until all nodes have set their transmission distances (the number of such update cycles is bounded by $D_M/\delta$, where $D_M$ takes the value of the initial (maximum) transmission distance. The nature of the algorithm ensures that the overall normalized cost $C = \Sigma\, D_i^2/D_o^2$ is equal or better than that of the uniform distance network while ensuring that $L < h$. Moreover, it is guaranteed that no node can reduce its transmission distance without violating this condition.

We measure the efficacy of adaptive networks with respect to uniform distance networks using eight different nodal topologies for 256 nodes (Fig. 2, coordinate origin (0, 0) is the bottom-left corner of each panel) namely: A) random; B) random in three 200 x 200 clusters centered at x-y coordinates (100, 100), (300, 400), (500, 200) with 50 nodes per cluster, and the remaining nodes randomly distributed over the 600 x 600 grid; C) 60% within a 200-radius central cluster with coordinates ($\rho\cos\theta$, $\rho\sin\theta$), where $\rho$ is randomly generated in the range (0, 200) and $\theta$ in the range of (0, $2\pi$), 40% randomly distributed;

D) star configuration (five 200 x 200 randomly-distributed clusters centered at x-y coordinates (100, 100), (100, 500), (500, 500), (500, 100), and (300, 300) with 50 nodes per cluster except the central cluster with 56 nodes; E) uniform lattice; F) radial with coordinates ($k \cos[2\pi k(k+1)/96] + 300$ m, $k \sin[2\pi k(k+1)/96] + 300$) and $k$ is an integer from 1 to 256; G) distributed along preset lines; and H) random walk starting at the center. These configurations were chosen to represent a variety of geographical or nodal deployment constraints.

Figure 3 compares the network average path length ($L$) and cost ($C$) for the adaptive (circles) and uniform radii (square) methods for each node topology for varying $h$ (5 to 30 hops in 5-hop increments). In general, the adaptive network provides significant cost savings given the same $h$ over the uniform distance network. More specifically, we observe the following trends:

a) For larger values of $h$, the largest nearest neighbor distance sets the minimum value of $D$. Figure 3c shows that for $h > 5$, the uniform radii method reached its limit due to the presence of a single relatively isolated node (Fig. 2c, bottom center). In general, $L$ increases with $h$.

b) The uniform distance and adaptive network solutions converge to the same value for large values of $h$ if and only if a constant nearest neighbor distance exists. In Fig. 3e, convergence is achieved at $h = 30$ hops: the optimum distance is the minimum node-to-node distance and 30 hops exactly cover the distance from corner-to-corner in the uniform grid. It is worth noting that the line topology (Fig. 2g) would have belonged to this category had all the lines been joined together. In practical situations, such a break in the line may have been caused by a few nonfunctional nodes at critical junctions and highlights the strength of adaptive networks in allocating increased power output only at the boundary nodes.

c) In regions of high density, the adaptive method significantly reduces node transmission when subject to the $h$ constraint. The significant cost savings can alternatively be used to reduce $L$ for the same total cost (Fig 3).

We may take $h$ to represent transmission lifetime and the number of signals that are being transmitted and received by a node increases with this lifetime. Taking $h$ as a proxy for network load, we compare the performance of the adaptive and uniform distance networks using two metrics: a) relative cost $\kappa = C_A / C_U$; and b) length factor $\lambda = L_A/L_U$. Figure 4 shows the tradeoff between $\kappa$ and $\lambda$ for different topologies and $h$ values. For this study we considered both $\alpha = 2$ and $\alpha = 4$. Topology-dependent effects are particularly evident for: a) Fig. 2c topology, where increasing $h$ results in a marginal improvement of cost but results in a significant length factor change; and b) Fig. 2e topology, where the uniform grid creates unique stepwise relationships between $h$ and $D$. In all other cases where a smooth distribution of the nearest neighbor distances exists, we obtain $\kappa \sim 0.2$ which translates to about 80% cost savings at the expense of a two- or three-fold length factor change under the same network load ($h$). In some cases the improvement exceed 90%. Slightly more substantial improvements are obtained when $\alpha = 4$ (Fig. 4, open symbols).

For a constant signal transmission rate, the number of incoming links (in-degree) [42] to a node is an indication of the congestion at the node. The relative mean in-degree of the adaptive to the uniform distance network $\gamma$ (Fig. 5) is significantly less than unity ($0.1 < \gamma < 0.6$ except when $h = 30$ for the grid network). For a broadcasting network, a drop in the number of incoming links is accompanied by a corresponding reduction in the likelihood of signal interference. A system using such a network needs fewer independent channels i.e. fewer distinct chemicals or transmission frequencies.

As mentioned in the introduction, our algorithm can thus also be considered as a member of this class of "congestion-aware protocols" [6, 11, 14-17, 20]. Methods of congestion reduction include message-specific retransmission (e.g. listening and transmitting within a limited frequency range, responding only to specific chemical transmitters, not relaying messages from certain nodes); or by selectively inserting absorbing material (e.g. cellular matrix, absorbing walls) to inhibit transmission between specific nodes that are proximate.

Our method effectively changes the link statistics of a network without changing the node distribution. If the node distribution constraint is lifted, we expect the emergence of scale-free networks [43].

The scaling (and scalability) of our results to larger system sizes can be analyzed from three perspectives: 1) the maximum number of hops needed to cross the system; 2) the characteristic power requirement on a node; and 3) the time it takes for the recipient to receive a message. For a network with a largely uniform density of nodes and ignoring corrections to scaling due to fluctuations, we can analyze the analogous system of $N$ discs of radius $r$ in a grid with side $s$, with a network density $\rho = N/s^2$. First, for fixed $r$ the maximum number of hops ($h$) needed to cross the system increases linearly in each linear dimension, $h \sim s/r$, it therefore grows as the square root of the density or network size, i.e. $h \sim s\sqrt{\rho} = \sqrt{N}$. Second, since the characteristic cost of a node is proportional to its transmission area, $P = r^2$, the characteristic cost decreases with increasing density: $P \sim s^2/N = 1/\rho$. Finally, the time it takes for a recipient to receive a message depends on the messaging requirements of the network. If communication is global so that there is equal probability for any two nodes to communicate, the required time scales with the maximum number of hops and thus increases with linear dimension and the square root of density $t \sim h \sim s\sqrt{\rho}$. On the other hand, if it is more likely for local nodes to communicate, the time is independent of system size.

## V. Conclusions

We have shown that an adaptive node transmission distance strategy of preferential detachment, given fixed node locations and fixed signal lifetime, trades off lower cost and efficient channel usage (lower number of incoming links) with small number of hops. Such a strategy provides an average of 80% and as much as 90% transmission cost savings over a uniform node transmission distance network along with a 2 to 3-fold increase in the average number of hops in a transmission and a 40% to 90% drop in the mean number of incoming links per node. This tradeoff is achieved with a fixed upper bound on the path length.

**Acknowledgments**

We thank R. Cohen for his comments. This work was supported in part by Sandia National Laboratories under US DoE Contract DE-AC04-94AL85000.

**Captions to Figures**

Fig. 1. **a**, Uniform distance network of 256 nodes (circles) in a 600 x 600 grid with transmission distance $D = 60$. Links are shown where inter-node distance is one hop ($d_{ij} = 1$). **b**, Variation of the average path length $L$ and reachable pairs $R$ for the network in **a**, $h = 20$ hops.

Fig. 2. Test nodal distributions: **a,** random; **b,** random in three clusters; **c,** 60% within a 200-radius central cluster, 40% randomly distributed; **d,** star configuration; **e,** uniform lattice; **f,** radial; **g,** lines; and **h,** random walk starting at the center.

Fig. 3. Panels correspond to node distributions (Fig 2) for adaptive (circles) and uniform radii (squares) methods. Points indicate the minimum possible cost $C$ and average path length $L$ with $h = 5, 10, 15, 20, 25, 30$. In general, $L$ increases with $h$.

Fig. 4. Relative power consumption $\kappa = C_A/C_U$ and relative path length $\lambda = L_A/L_U$ of adaptive networks with respect to uniform distance networks. Power consumption is calculated by summing the values of $D_i^\alpha$ when $\alpha = 2$ (filled symbols) and $\alpha = 4$ (open symbols).

Fig. 5. Congestion at the node level drops with the use of an adaptive network as shown by the variation of the mean in-degree ratio $\gamma = I_A/I_U$ with the maximum packet lifetime, where $I_A$ and $I_U$ are the mean number of incoming links to a node for the adaptive and uniform distance networks, respectively.

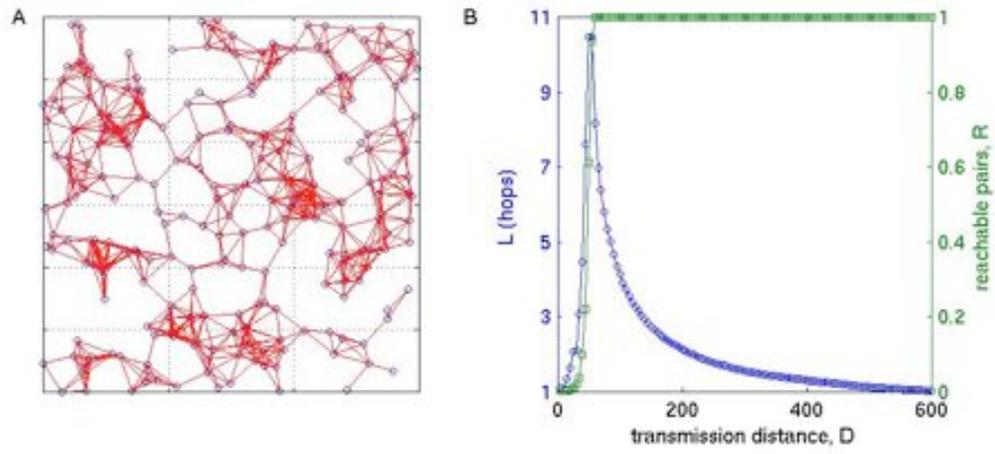

Figure 1.

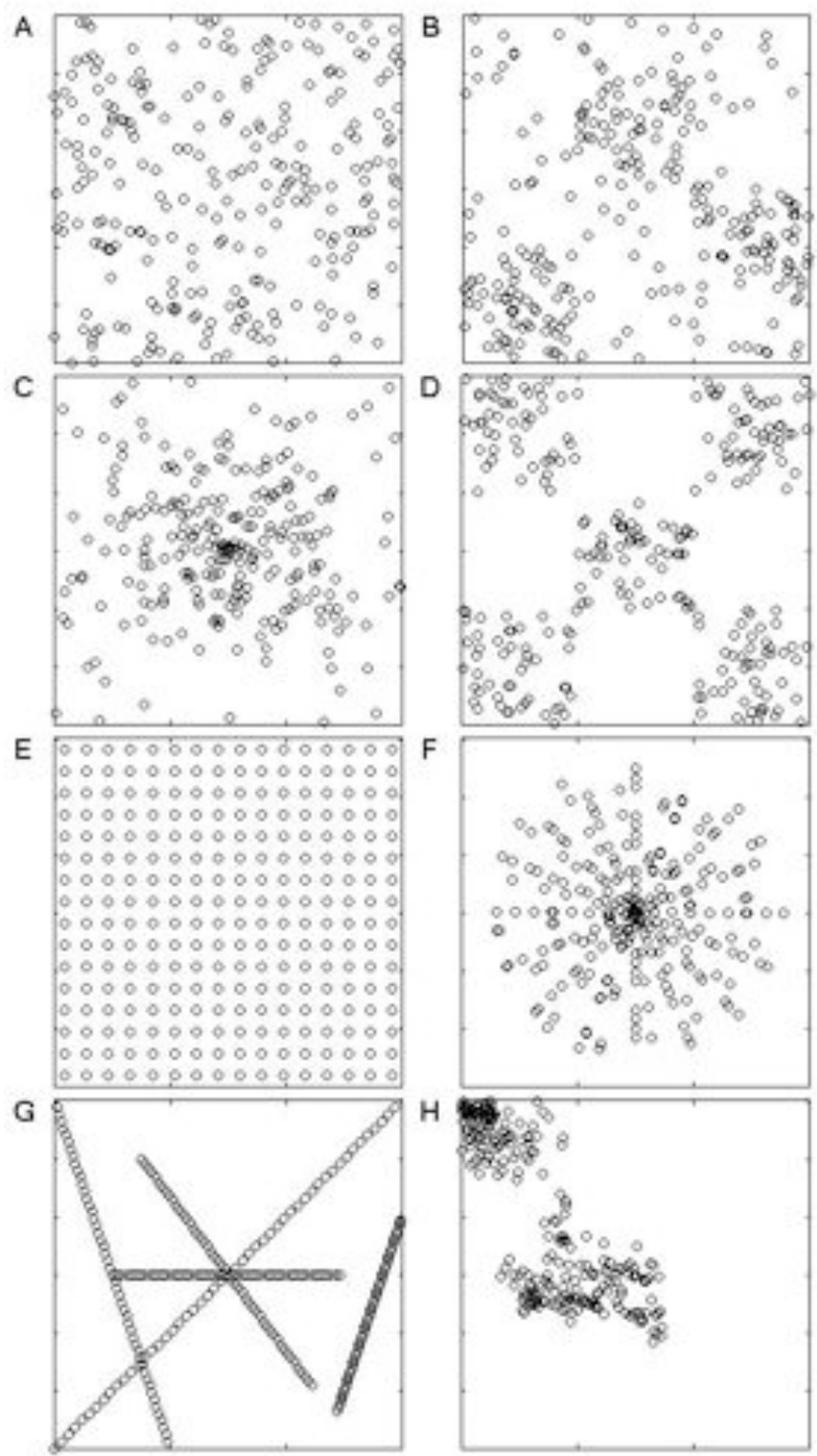

Figure 2.

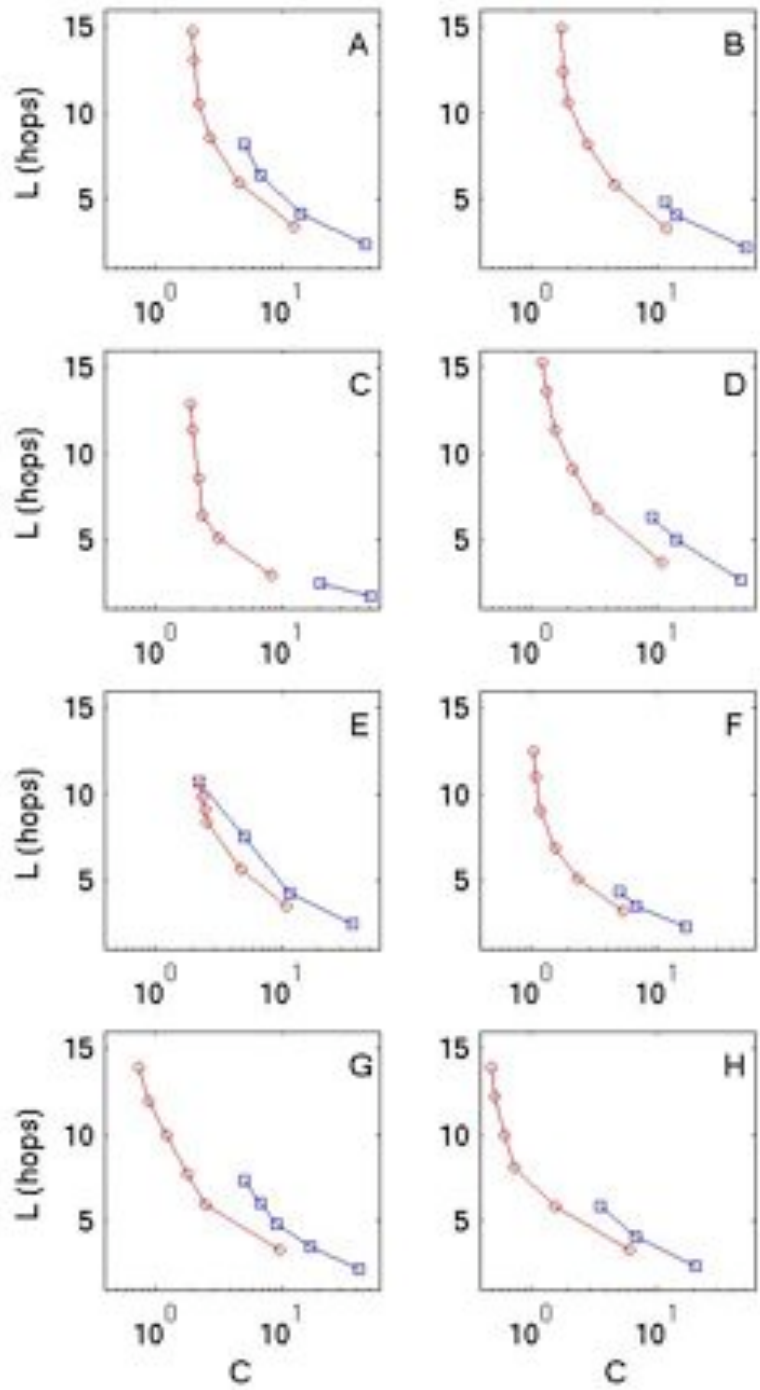

Figure 3.

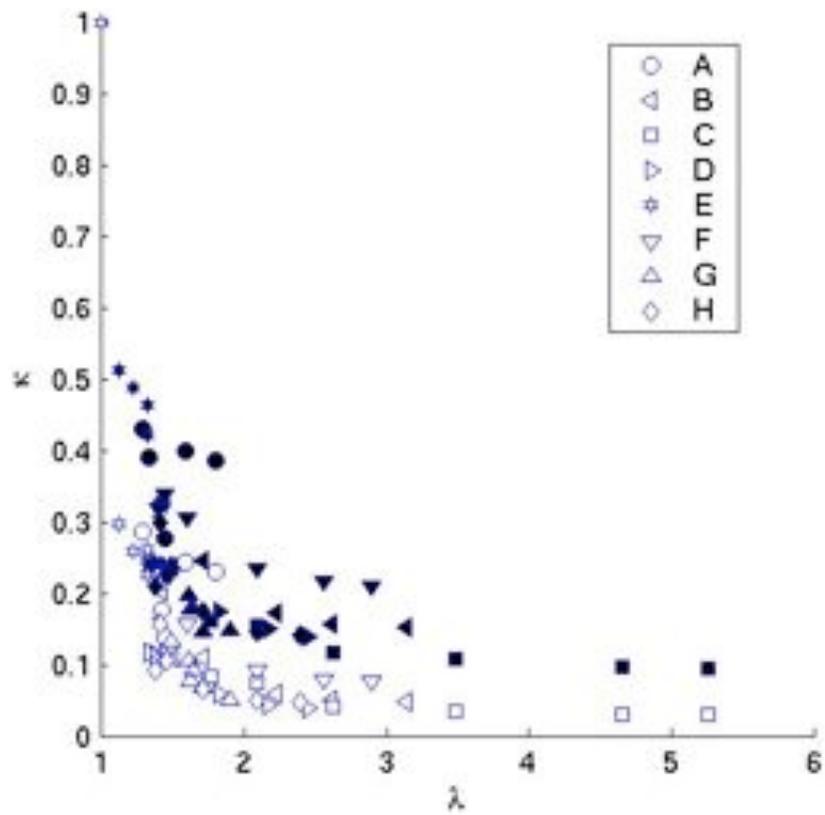

Figure 4.

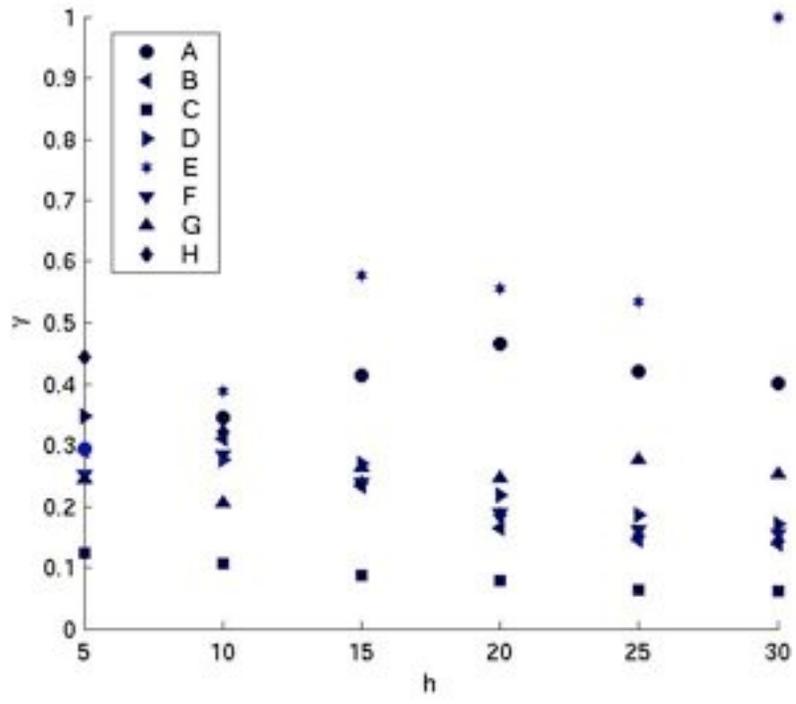

Figure 5.